\documentclass[twocolumn,pre,superscriptaddress,showpacs]{revtex4}

\usepackage{graphicx,amsmath}

\newcommand{\al}{\alpha}

\begin{document}

\title{Inequivalence of time and ensemble averages in ergodic systems:
exponential versus power-law relaxation in confinement}

\author{Jae-Hyung Jeon}
\email{jeonjh@gmail.com}
\affiliation{Department of Physics, Tampere University of Technology, FI-33101
Tampere, Finland}
\author{Ralf Metzler}
\email{rmetzler@uni-potsdam.de}
\affiliation{Institute for Physics \& Astronomy, University of Potsdam,
D-14476 Potsdam-Golm, Germany}
\affiliation{Department of Physics, Tampere University of Technology, FI-33101
Tampere, Finland}

\begin{abstract}
Single particle tracking has become a standard tool to investigate diffusive
properties, especially in small systems such as biological cells. Usually
the resulting time series are analyzed in terms of time averages over
individual trajectories. Here we study confined normal as well as anomalous
diffusion modeled by fractional Brownian motion and the fractional Langevin
equation, and show that even for such ergodic systems time-averaged quantities
behave differently from their ensemble averaged counterparts, irrespective of
how long the measurement time becomes. Knowledge of the exact behavior of
time averages is therefore fundamental for the proper physical interpretation
of measured time series, in particular, for extraction of the relaxation time
scale from data.
\end{abstract}


\maketitle

\section{Introduction}

Due to recent advances in single particle tracking techniques, analyses based
on single trajectory averages have been widely employed to study diffusion in
complex systems, e.g., of large biomolecules and tracers in living cells
\cite{braeuchle}. Examples include the motion in the cellular cytoplasm of 
messenger RNA molecules \cite{golding}, chromosomal loci \cite{weber}, lipid
granules \cite{jh}, , and viruses \cite{georg}, telomeres in cell nuclei
\cite{garini}, or of protein channels in the
cell membrane \cite{weigel}. Under the assumption of ergodicity, i.e., the
equivalence of (long) time averages (TA) with ensemble averages (EA), the
physical interpretation is often based on the time series analysis of single 
trajectories. For instance, particle-to-particle diffusion properties are
typically studied via TA mean squared displacements (MSD) of individual
time series $x(t)$,
\begin{equation}
\label{tamsddef}
\overline{\delta^2(\Delta)}=\frac{1}{T-\Delta}\int_0^{T-\Delta}\Big[x(t+\Delta)-
x(t)\Big]^2dt,
\end{equation}
where $\Delta$ is the lag time and $T$ the length of the time series. Invoking
ergodicity arguments, it is tacitly assumed that $\overline{\delta^2(\Delta)}$
corresponds to the EA MSD $\langle x^2(t)\rangle$ with the identification $t
\leftrightarrow\Delta$, in the limit of long measurement times (i.e., $T\to
\infty$). For \emph{free\/} normal diffusion one can indeed show analytically
that $\langle x^2(t)\rangle=\overline{\delta^2(t)}=2K_1t$ as $T\to\infty$
\cite{He,lubelski}. At finite $T$, the result for $\overline{\delta^2(\Delta)}$
will generally show trajectory-to-trajectory variations. However, a similar
equivalence still holds when $\overline{\delta^2}$ is averaged over many
individual
trajectories: $\langle x^2(t)\rangle=\langle\overline{\delta^2(t)}\rangle$
\cite{He}. In what follows we use the symbol $\overline{\delta^2}$
when $T\to\infty$, and $\langle\overline{\delta^2}\rangle$ for finite $T$,
unless specified otherwise.
For anomalous diffusion of the form $\langle x^2(t)\rangle=2K_{
\alpha}t^{\alpha}$ with anomalous diffusion constant $K_{\alpha}$ of physical
dimension $\mathrm{cm}^2/\mathrm{sec}^{\alpha}$ and anomalous diffusion exponent
$\alpha$ ($0<\alpha<2$) \cite{report}, the same conclusion holds if the process
is described by fractional Brownian motion (FBM) or the fractional Langevin
equation (FLE) \cite{deng,stas2,goychuk}.

In contrast, disagreements between TA and EA are not surprising for non-ergodic
processes. A prominent example is anomalous diffusion described by continuous
time random walks (CTRW) with diverging characteristic waiting times
\cite{montroll,scher,web}: while the EA MSD scales as $\langle x^2(t)\rangle\simeq
t^{\kappa}$ with $0<\kappa<1$, the TA MSD grows \emph{linearly\/} with the lag
time, $\overline{\delta^2(\Delta)}\simeq\Delta$ for free motion
\cite{He,lubelski}. Under confinement one observes $\overline{\delta^2(\Delta)}
\simeq\Delta^{1-\kappa}$ instead of the saturation plateau of the EA
\cite{stas2,stas3,neusius}. Recently it was found that the TA MSD of tracers
in living cells indeed exhibit such CTRW behavior \cite{jh,weigel}.

Here, we show that \emph{even for ergodic processes\/} the TA may differ from
the EA. This a priori unexpected discrepancy arises from the fact that generally
dynamic variables are not well-defined in the TA sense, and therefore care is
necessary when interpreting TA based on knowledge about the corresponding EA.
We explicitly study this effect for stochastic processes of regular Brownian,
FBM, and FLE types, confined in an harmonic potential. Processes of the FBM and
FLE kind are closely associated with the motion of tracer molecules in viscous
environments, single file diffusion, monomer motion in polymers, or the relative
motion of aminoacids in proteins \cite{singlefile}. They have also been
identified as stochastic mechanisms for the tracer motion in living cells and
reconstituted crowding systems \cite{weiss,weber,jh,marcin}.

\begin{figure}
\includegraphics[width=8cm]{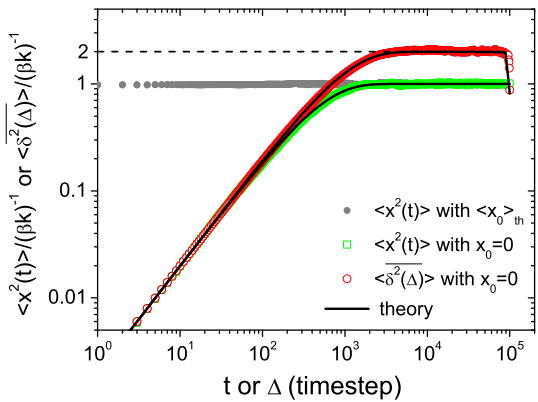}
\caption{EA and TA MSD for a Brownian particle in an harmonic potential. Solid
lines: Eqs.~\eqref{ea}, \eqref{ta}. Symbols: simulations with
$\gamma=1$, $\beta=1$, $k=1$, time step $\delta t=0.001$, and measurement time
$T=10^5$. The flat curve corresponds to thermal initial conditions.}
\label{brownian_har}
\end{figure}

Consider first an overdamped Brownian particle in the harmonic potential
$U(x)=kx^2/2$ of stiffness $k$. With initial position $x(0)=0$ the EA MSD is
\begin{equation}
\label{ea}
\langle x^2(t)\rangle=(1-e^{-2kt/\gamma})/[\beta k],
\end{equation}
while the EA taken over the TA MSD \eqref{tamsddef} becomes
\begin{eqnarray}
\nonumber
\left<\overline{\delta^2(\Delta,T)}\right>&=&\frac{2}{\beta k}\left(1-e^{-k
\Delta/\gamma}\right)\\
&&\hspace*{-2.2cm}
+\frac{\gamma}{2k(T-\Delta)}\left(e^{k\Delta/\gamma}-1\right)^2
\left(e^{-2kT/\gamma}-e^{-2k\Delta/\gamma}\right).
\label{ta}
\end{eqnarray}
Here $\gamma$ is the friction coefficient and $\beta$ the Boltzmann factor.
Both quantities are identical initially, before confinement effects come
into play: $\langle x^2(t)\rangle\sim2K_1t\sim\langle\overline{\delta^2(t)}
\rangle$ with $K_1=1/[\beta\gamma]$. However, for $T,\Delta\rightarrow\infty$
(with $T-\Delta\to\infty$) the TA MSD
converges to $\overline{\delta^2(\Delta)}\rightarrow2/[\beta k]$, \emph{twice\/} the thermal value $\langle x^2\rangle_{\mathrm{th}}=1/[\beta k]$. The
difference between TA and EA is shown in Fig.~\ref{brownian_har} for both the
analytical results and simulations, with excellent agreement. Note the sudden
dip of the TA MSD at $\Delta\approx T$ at the finite measurement time $T$, at
which the limiting behavior $\langle\overline{\delta^2(\Delta\rightarrow T)}
\rangle=\langle x^2(T)\rangle$ is observed. These features are generic for the
definition of the TA MSD (\ref{tamsddef}) under confinement, compare
Ref.~\cite{stas2,stas3}.

What results will be obtained for more complicated, non-Brownian motion? We
analyze the case of anomalous diffusion governed by FBM and FLE and show that
the entire relaxation dynamics is significantly different for the TA, despite
the ergodic nature of these processes. Knowledge
of the exact behavior of TA quantities is imperative for the correct
physical interpretation of time series, in particular, to extract the
relaxation time.

\section{Fractional Brownian motion}

FBM $x_\al(t)$ in an
external potential $U$ follows the Langevin equation
\begin{equation}
\label{lang}
\frac{dx_\al(t)}{dt}=-kx_\al(t)+\xi_\al(t),
\end{equation}
driven by fractional Gaussian noise $\xi_\alpha(t)$ of zero mean
$\langle \xi_\al(t)\rangle=0$ and slowly decaying, power-law autocorrelation
($t\neq t'$) \cite{mandelbrot,olekssi}
\begin{equation}
\label{autofGn}
\langle \xi_\al(t)\xi_\al(t')\rangle\simeq\alpha K_{\alpha}(\alpha-1)|t-t'|^{
\alpha-2}.
\end{equation}
In free space, $\langle x_\al^2(t)\rangle=2K_{\alpha}t^{\alpha}$ \cite{deng}.
Note the change of sign in Eq.~(\ref{autofGn}) between antipersistent
subdiffusion $0<\alpha<1$ and persistent superdiffusion $1<\alpha<2$.
Different to subdiffusive CTRW processes with diverging waiting time scale,
FBM does not exhibit ageing effects. In fact, the free space propagator is
the Gaussian \cite{lutz}
\begin{equation}
\label{prop}
P(x,t)=\sqrt{\frac{1}{4\pi K_{\alpha}t^{\alpha}}}\exp\left(-\frac{x^2}{4K_{
\alpha}t^{\alpha}}\right),
\end{equation}
whose smooth shape contrasts the pronounced cusps at the initial position
in subdiffusive CTRW processes \cite{scher,report}. Moreover, the propagator
(\ref{prop}) obeys a time-local diffusion equation with time-dependent
diffusivity \cite{lutz}.

\begin{figure}[tb]
\includegraphics[width=8cm]{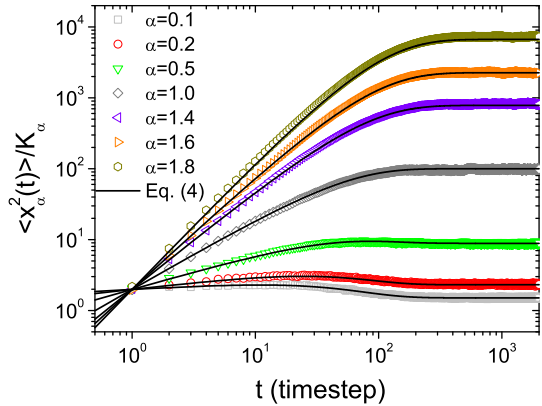}
\caption{EA MSD $\langle x_\al^2(t)\rangle$ for FBM in an harmonic potential.
Solid lines: Eq.~\eqref{msdfbm}. Symbols: simulations with
parameters $k=0.01$, $T=2048$, and $x_0=0$.}
\label{Ffbm_har}
\end{figure}

The formal solution of the FBM-Langevin equation (\ref{lang}),
\begin{equation}
x_\al(t)=\int_0^te^{-k(t-t')}\xi_\al(t')dt',
\end{equation}
and Eq.~(\ref{autofGn}) yield the position autocorrelation function
\begin{eqnarray}
\nonumber
\langle x_\alpha(t_1)x_\alpha(t_2)\rangle&=&K_\alpha\left\{e^{-kt_1}t_2^{
\alpha}+e^{-kt_2}t_1^{\alpha}-|t_2-t_1|^{\alpha}\right\}\\
\nonumber
&&\hspace*{-2.0cm}
+\frac{K_\alpha}{2k^{\alpha}}\left\{e^{-k|t_2-t_1|}\gamma(\alpha+1,kt_1)\right.\\
\nonumber
&&+e^{k|t_2-t_1|}\gamma(\alpha+1,kt_2)\\
\nonumber
&&\left.-e^{k|t_2-t_1|}\gamma(\alpha+1,k|t_2-t_1|)\right\}\\
\nonumber
&&\hspace*{-2.0cm}
+\frac{kK_\alpha}{2(\alpha+1)}|t_2-t_1|^{\alpha+1}e^{-k|t_2-t_1|}\\
\nonumber
&&\times M(\alpha+1;\alpha+2;k|t_2-t_1|)\\
\nonumber
&&\hspace*{-2.0cm}
-\frac{kK_\alpha}{2(\alpha+1)}t_1^{\alpha+1}e^{-k(t_1+t_2)}M(\alpha+1;\alpha
+2;kt_1)\\
&&\hspace*{-2.0cm}
-\frac{kK_\alpha}{2(\alpha+1)}t_2^{\alpha+1}e^{-k(t_1+t_2)}M(\alpha+1;\alpha
+2;kt_2).
\end{eqnarray}
For the EA MSD we then find
\begin{eqnarray}
\langle
x_\al^2(t)\rangle&=&\frac{K_{\alpha}}{k^{\alpha}}\gamma(\alpha+1,kt)+2
K_{\alpha}t^{\alpha}e^{-kt}\nonumber
\\
&&-\frac{kK_{\alpha}}{\alpha+1}t^{\alpha+1}e^{-2kt}M(\alpha+1;\alpha+2;kt),
\label{msdfbm}
\end{eqnarray}
where $\gamma(z,x)=\int_0^x dt e^{-t}t^{z-1}$ is the incomplete $\gamma$
function and
\begin{equation}
M(a;b;z)=\frac{\Gamma(b)}{\Gamma(b-a)\Gamma(a)}\int_0^1e^{zt}t^{a-1}
(1-t)^{b-a-1}dt
\end{equation}
is the Kummer function~\cite{abramowitz}. Fig.~\ref{Ffbm_har}
shows simulations of FBM in an harmonic potential for various $\alpha$ values,
demonstrating excellent agreement with result \eqref{msdfbm}. Asymptotic
expansion of Eq.~\eqref{msdfbm} at short times $t\ll k^{-1}$ yields free
anomalous diffusion $\langle x_\al^2(t)\rangle\sim2K_{\alpha}t^{\alpha}$.
Close to stationarity, we find
\begin{equation}
\langle x_\al^2(t)\rangle\sim\langle x_\al^2\rangle_{\mathrm{th}}-\frac{2}{k^2}
\alpha(\alpha-1)K_{\alpha}t^{\alpha-2}e^{-kt},
\label{msdexpansion}
\end{equation}
exponentially approaching the stationary value
\begin{equation}
\langle x_\al^2\rangle_{\mathrm{th}}=\frac{K_{\alpha}}{k^{\alpha}}\Gamma(
\alpha+1)
\label{fbmst}
\end{equation}
with the single characteristic time scale $k^{-1}$ in the exponential function:
as observed in Fig.~\ref{Ffbm_har}, beyond $t>k^{-1}$ the stationary state is
attained independent of $\alpha$. This property enables one
to study the confinement effect by analyzing the relaxation of $\langle
x_\al^2(t)\rangle$. We also note an interesting feature of the relaxation
dynamics in the intermediate timescale: somewhat counterintuitively the
subdiffusive particle overshoots $\langle x_\al^2\rangle_{\mathrm{th}}$ before
a depression back to this value, while for superdiffusion we observe a monotonic
increase (the sign of the second term in Eq.~\eqref{msdexpansion} depends
on $\alpha$). Note that the $\alpha$-dependence of the plateau value
(\ref{fbmst}), a reminder of the fact that FBM is driven by an external noise
and thus not subject to the fluctuation dissipation theorem, in contrast to
FLE motion discussed below. Phenomenologically, both processes are very
similar.

For the TA MSD in the limit $T\rightarrow\infty$ we obtain the expression
\begin{eqnarray}
\nonumber
\overline{\delta^2(\Delta)}&=&2K_{\alpha}\Gamma(\alpha+1)/k^{\alpha}+2K_{\alpha}
\Delta^{\alpha}\\
\nonumber
&&\hspace*{-1.2cm}
-\frac{K_{\alpha}}{k^\alpha}\left\{e^{k\Delta}\Gamma(\alpha+1,k\Delta)+e^{-k\Delta}\Gamma(
\alpha+1)\right\}\\
&&\hspace*{-1.2cm}
-\frac{kK_{\alpha}}{\alpha+1}\Delta^{\alpha+1}e^{-k\Delta}M(\alpha+1;\alpha+2;k\Delta),
\label{tamsdfbmhar}
\end{eqnarray}
where $\Gamma(z,x)=\int_x^\infty dt e^{-t}t^{z-1}$ is the complementary
incomplete $\gamma$ function. Comparison with the EA MSD,
Eq.~\eqref{msdfbm}, demonstrates a completely different functional behavior
\emph{over all time scales}, except in the short-time limit for which
confinement is negligible. In particular, at $\Delta\rightarrow\infty$, we find
$\overline{\delta^2(\Delta)}=2\langle x_\al^2\rangle_{\mathrm{th}}$ for all
$\alpha$.

\begin{figure}
\includegraphics[width=8cm]{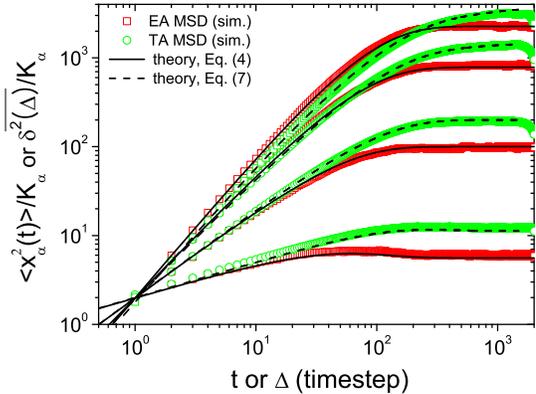}
\caption{EA and TA MSD, $\langle x_\al^2(t)\rangle/K_\al$ and $\overline{
\delta^2(\Delta)}/K_\al$ for FBM in an harmonic potential ($\alpha=0.40$,
1.0, 1.4, 1.60, bottom to top). Solid and dashed lines: analytical results
(\ref{msdfbm}) and (\ref{tamsdfbmhar}). Symbols: simulations. Parameters as
in Fig.~\ref{Ffbm_har}.}
\label{Ftamsdfbm2}
\end{figure}

The fundamental difference of the relaxation dynamics of $\overline{\delta^2(
\Delta)}$ and $\langle x_\al^2(t)\rangle$ is evidenced in Fig.~\ref{Ftamsdfbm2}, in excellent agreement with Eqs.~(\ref{msdfbm}) and (\ref{tamsdfbmhar}).
In contrast to the exponential relaxation of Eq.~(\ref{msdfbm}), the TA MSD
shows a power-law approach to the limiting value $2\langle x_\al^2\rangle
_{\mathrm{th}}$, except for the Brownian limit $\alpha=1$. This
is manifested in the asymptotic form of $\overline{\delta^2(\Delta)}$
at $\Delta\rightarrow\infty$,
\begin{eqnarray}
\nonumber
\overline{\delta^2(\Delta)}&\sim&2\langle x_\al^2\rangle_{\mathrm{th}}
-\frac{K_{\alpha}\Gamma(\alpha+1)}{k^2}e^{-k\Delta}\\
&&-\frac{2\alpha(\alpha-1)K_{\alpha}}{k^2\Delta^{2-\alpha}}.
\label{tamsdfbmexpansion}
\end{eqnarray}
The transient second term becomes the leading order at $\alpha=1$.
Surprisingly, in Eq.~\eqref{tamsdfbmexpansion} the relaxation dynamics is
determined by the power exponent $\alpha-2$. Moreover, no characteristic
time scale exists beyond which the MSD could be regarded saturated. For
subdiffusion, as the algebraic decay is relatively fast ($\sim\Delta^{-\kappa}$
with $1<\kappa<2$), the MSD appears saturated at sufficiently long measurement
time. However, the superdiffusive MSD relaxes very slowly as $\alpha$ is closer
to 2 ($\sim\Delta^{-\kappa}$ with $0<\kappa<1$). Due to this the corresponding
MSD does not show saturation even at long measurement time $T$. Only in the
limit $\Delta\to T$ the TA dips back to the plateau value of the EA. In typical
experiments, however, this feature is obscured by poor statistics, and thus the
relaxation time would likely be overestimated from the TA MSD.

\section{Fractional Langevin equation}. 

The FLE describes ergodic anomalous
diffusion and fulfills the fluctuation-dissipation theorem \cite{deng}. In
the potential $U$, the FLE motion $y_\al(t)$ follows the dynamic equation
\cite{deng,jeon,kou}
\begin{eqnarray}
\nonumber
m\frac{d^2y_\al(t)}{dt^2}&=&-\gamma\int_{0}^t dt' |t-t'|^{\alpha-2}\frac{dy_\al}{
dt'}-ky_\al(t)\\
&&+\sqrt{\gamma/[\alpha(\alpha-1)\beta K_{\alpha}]}\xi_\al(t),
\label{glehar}
\end{eqnarray}
where $\xi_\al(t)$ represents fractional Gaussian noise, $m$ is the particle mass
and $\gamma$ the generalized friction coefficient. In the FLE, the dynamic
exponent of the noise is restricted to $1<\alpha<2$. This persistent noise
results in subdiffusive motion of the FLE in the overdamped limit. For unbiased
motion ($k=0$), $\langle y_\al^2(t)\rangle=\overline{\delta^2(t)}$ at $T\to
\infty$ \cite{deng}, and
\begin{equation}
\overline{\delta^2(\Delta)}=\frac{2\Delta^2}{\beta m}E_{\alpha,3}\left[-
\Gamma(\alpha-1)\frac{\gamma}{m}\Delta^{\alpha}\right],
\label{glefree}
\end{equation}
$E_{\alpha,3}(z)$ being a generalized Mittag-Leffler function. The latter
is defined via its Laplace image
\begin{equation}
\int_0^{\infty}e^{-ut}E_{\rho,\delta}\left(-\eta^*t^{\alpha}\right)=\frac{1}{
u^{\delta}+\eta^*u^{1-\alpha}}.
\end{equation}
In terms of a series expansion around $z=0$ and $z\to\infty$ this function
reads \cite{bateman}
\begin{equation}
E_{\rho,\delta}(z)=\sum_{n=0}^{\infty}\frac{z^n}{\Gamma(\delta+\rho n)}=
-\sum_{n=1}^{\infty}\frac{z^{-n}}{\Gamma(\delta-\rho n)}.
\end{equation}
The MSD (\ref{glefree}) accordingly turns from ballistic motion $\sim\Delta^2$
at short $\Delta$ to subdiffusion
$\sim\Delta^{2-\alpha}$ at long $\Delta$ \cite{jeon,kou}. In the presence of the
potential, the FLE (\ref{glehar}) can be solved analytically in the overdamped
limit, the stationary state yielding \cite{kou}
\begin{equation}
\langle y_\al(t_1)y_\al(t_2)\rangle_{\mathrm{th}}=\frac{1}{\beta k}E_{2-\alpha}
\left[-\frac{k}{\gamma\Gamma(\alpha-1)}|t_2-t_1|^{2-\alpha}\right],
\end{equation}
with $E_{2-\alpha}(z)=E_{2-\alpha,1}$. Thus, $\langle y_\al^2(t)\rangle$ has
the stationary value $\langle y_\al^2\rangle_{\mathrm{th}}=1/(\beta k)$
for any $\alpha$, contrasting the
$\alpha$-dependent result \eqref{fbmst} for FBM. Moreover for $T\rightarrow
\infty$ we obtain the TA MSD
\begin{equation}
\overline{\delta^2(\Delta)}=2\langle y_\al^2\rangle_{\mathrm{th}}\left(1-E_{2-
\alpha}\left[-\frac{k}{\gamma\Gamma(\alpha-1)}\Delta^{2-\alpha}\right]\right)
\label{tamsdglehar}
\end{equation}
for $\Delta\gtrsim\tau_c$, where the momentum relaxation time is \cite{jeon}
\begin{equation}
\tau_c=\left(m\frac{\Gamma(\alpha+3)}{2\Gamma(\alpha-1)(2^{\alpha+1}-1)\gamma}
\right)^{1/\alpha}.
\end{equation}
The TA MSD \eqref{tamsdglehar} behaves distinctly different from
its EA counterpart as well as the TA MSD \eqref{tamsdfbmhar} for FBM: the
TA MSD (\ref{tamsdglehar}) grows like $\sim\Delta^{2-\alpha}$ at intermediate
lag time, and eventually converges to $2\langle y_\al^2\rangle_{\mathrm{th}}$ for
all $\alpha$ as $\Delta\rightarrow\infty$. Similar to our above observations,
the long-time behavior of Eq.~(\ref{tamsdglehar}) exhibits a power-law
relaxation, namely,
\begin{equation}
\overline{\delta^2(\Delta)}\approx2\langle y_\al^2\rangle_{
\mathrm{th}}\left(1-\frac{\gamma}{k\Delta^{2-\alpha}}\right).
\end{equation}
As for FBM, the dynamic exponent of the TA MSD is independent of the confinement
($k$). Intriguingly the speed of convergence is slower as the driving noise
$\xi_\al$ is more persistent (i.e., when $\alpha\rightarrow2$). Therefore,
opposite
to the tendency shown in Fig.~\ref{Ftamsdfbm2} for FBM, in the FLE case the
\emph{slow\/} particle appears not to approach $2\langle y_\al^2\rangle_{
\mathrm{th}}$.

\begin{figure}
\includegraphics[width=8cm]{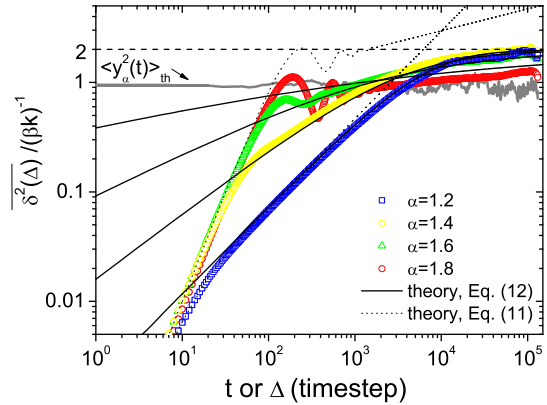}
\caption{TA MSD for FLE motion in an harmonic potential, for $\alpha=1.2$, 1.4,
1.6, and 1.8 (bottom to top), and $T=2^{17}$.
A representative EA MSD is included for $\alpha=
1.2$. Symbols: Simulations with time step $\delta t=0.001$, stiffness $k=100$,
mass $m=1$, friction coefficient $\gamma=100$, and $\beta=1$, with equilibrium
distribution of initial position $y_\al(0)$ and velocity $\dot{y}_\al(0)$. Full
lines: theoretical result \eqref{tamsdglehar}. Dotted lines: unbiased motion,
Eq. \eqref{glefree}, with momentum relaxation at $\alpha=1.4$ and 1.8,
illustrating ballistic scaling $\sim\Delta^2$ at $\Delta<\tau_c$.}
\label{Ftamsdglehar2}
\end{figure}

In Fig.~\ref{Ftamsdglehar2}, we further analyze FLE motion in an harmonic
potential in terms of the TA MSD for various $\alpha$. For times $\Delta\gtrsim
\tau_c$ and in the overdamped limit, the analytical form \eqref{tamsdglehar}
agrees well with the simulations results for all cases. While the TA MSD
approaches the thermal value $2\langle y_\al^2\rangle_{\mathrm{th}}$ algebraically,
in our simulation the slowest subdiffusive case (corresponding to $\alpha=1.8$)
does not show saturation even for long measurement times.
At times less than $\tau_c$
the TA MSD shows quadratic scaling $\overline{\delta^2}\simeq\Delta^2$.
The dynamics within this time range is explained well by the full solution
of the free space motion \eqref{glefree}, shown for $\alpha=1.2$ and 1.8. An
important feature resulting from this inertia effect are the oscillations of
$\overline{\delta^2}$ that are particularly pronounced as $\alpha$ approaches
2. These oscillations are intrinsic in the sense that they occur regardless of
the confinement, due to the strong persistence in $\xi_\al$ and inertia effects
\cite{stas}.

\section{Distribution of time averaged mean squared displacements}

\begin{figure*}
\includegraphics[width=8cm]{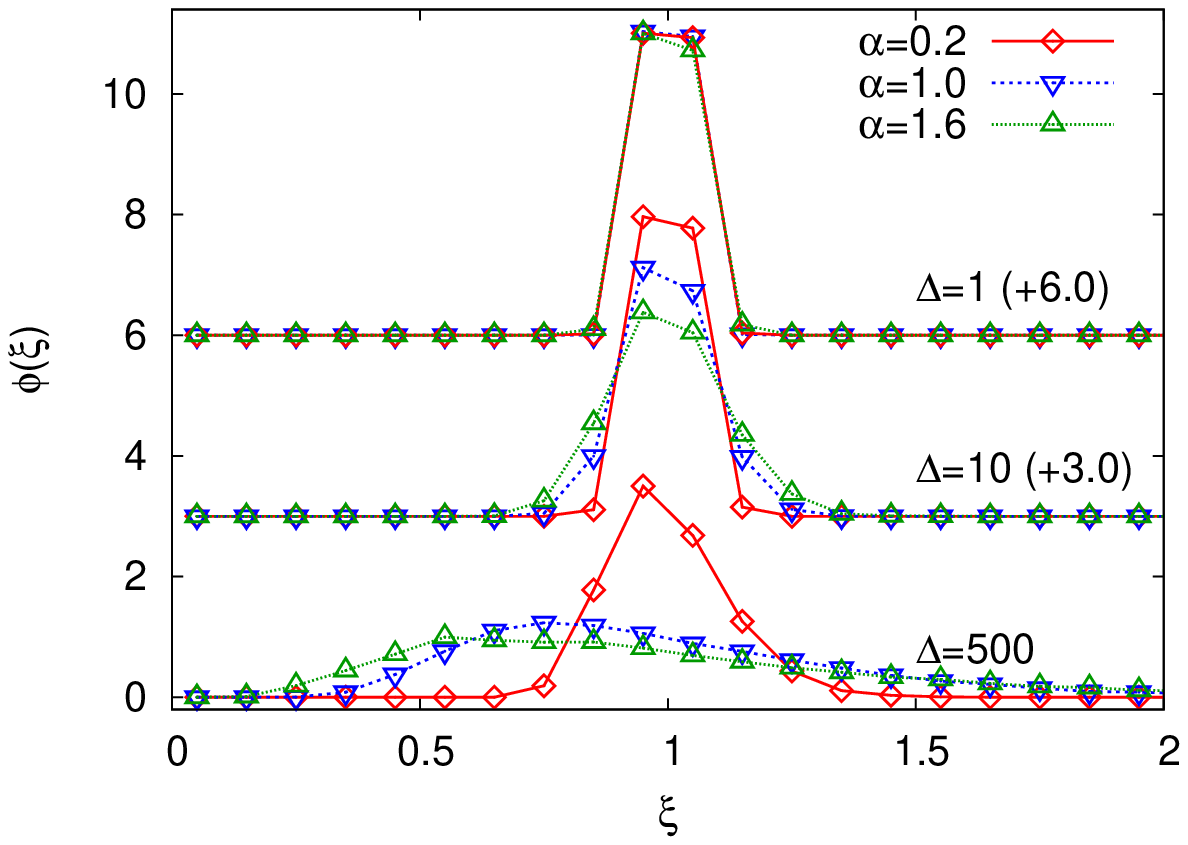}
\includegraphics[width=8cm]{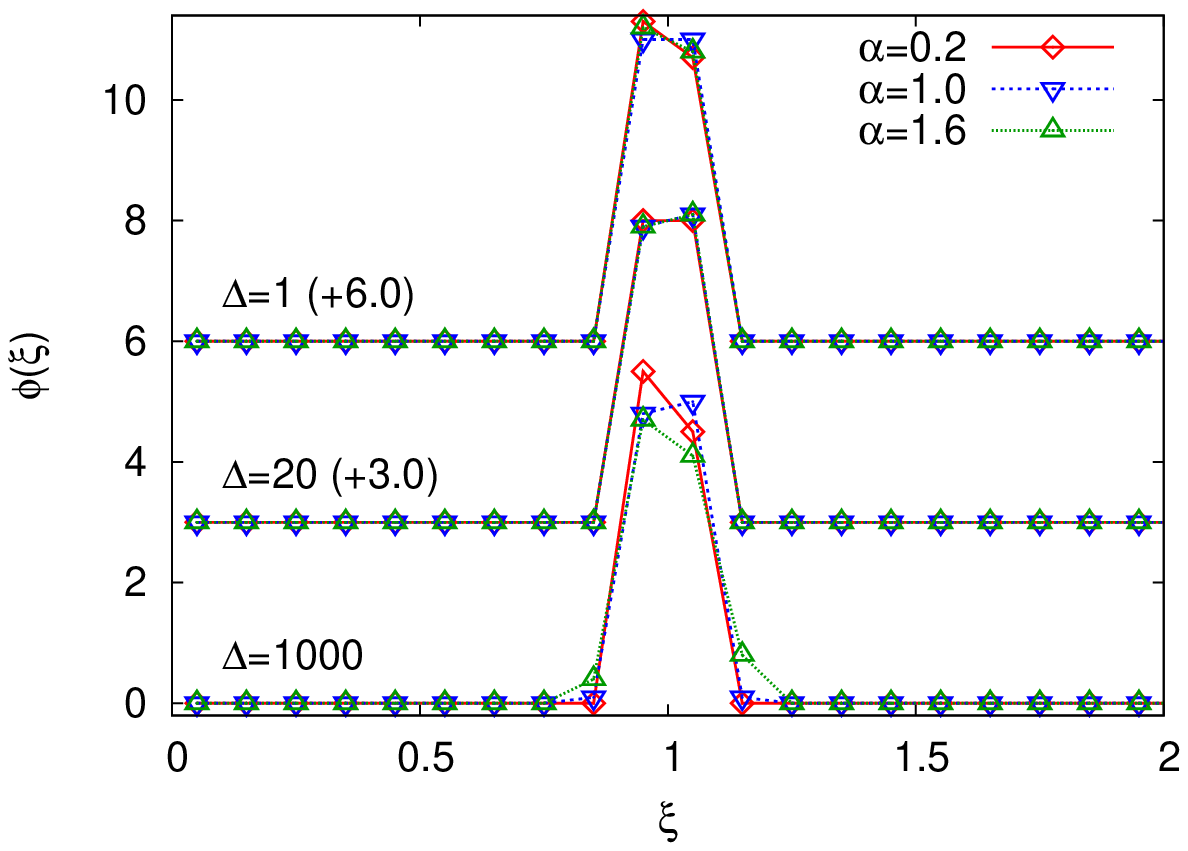}
\caption{Scatter distributions for fractional Brownian motion in an harmonic
potential. Left: $T=2^{11}=2,048$. Right: $T=2^{17}=13,1072$.
In each graph we compare the results for three different lag times.
The two upper sets of curves were shifted by the indicated amount,
for clarity.}
\label{sup1}
\end{figure*}

\begin{figure*}
\includegraphics[width=8cm]{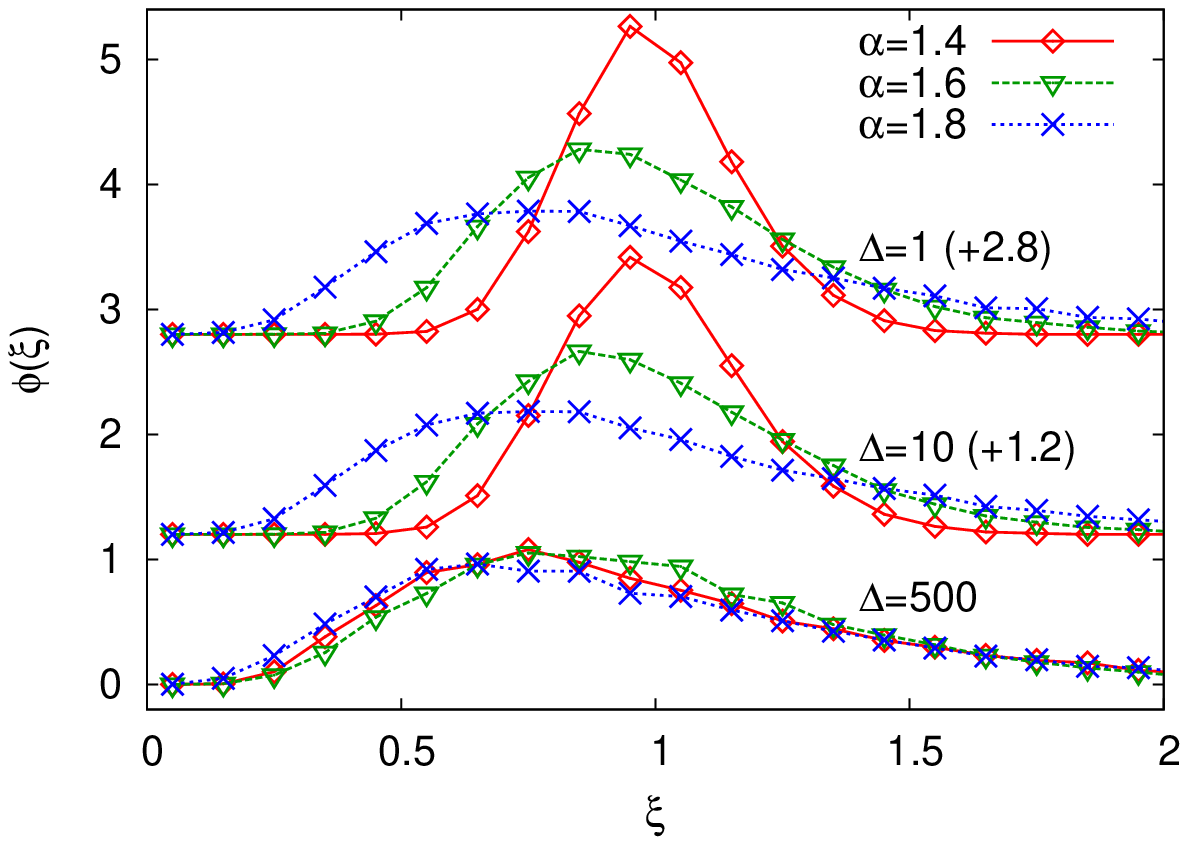}
\includegraphics[width=8cm]{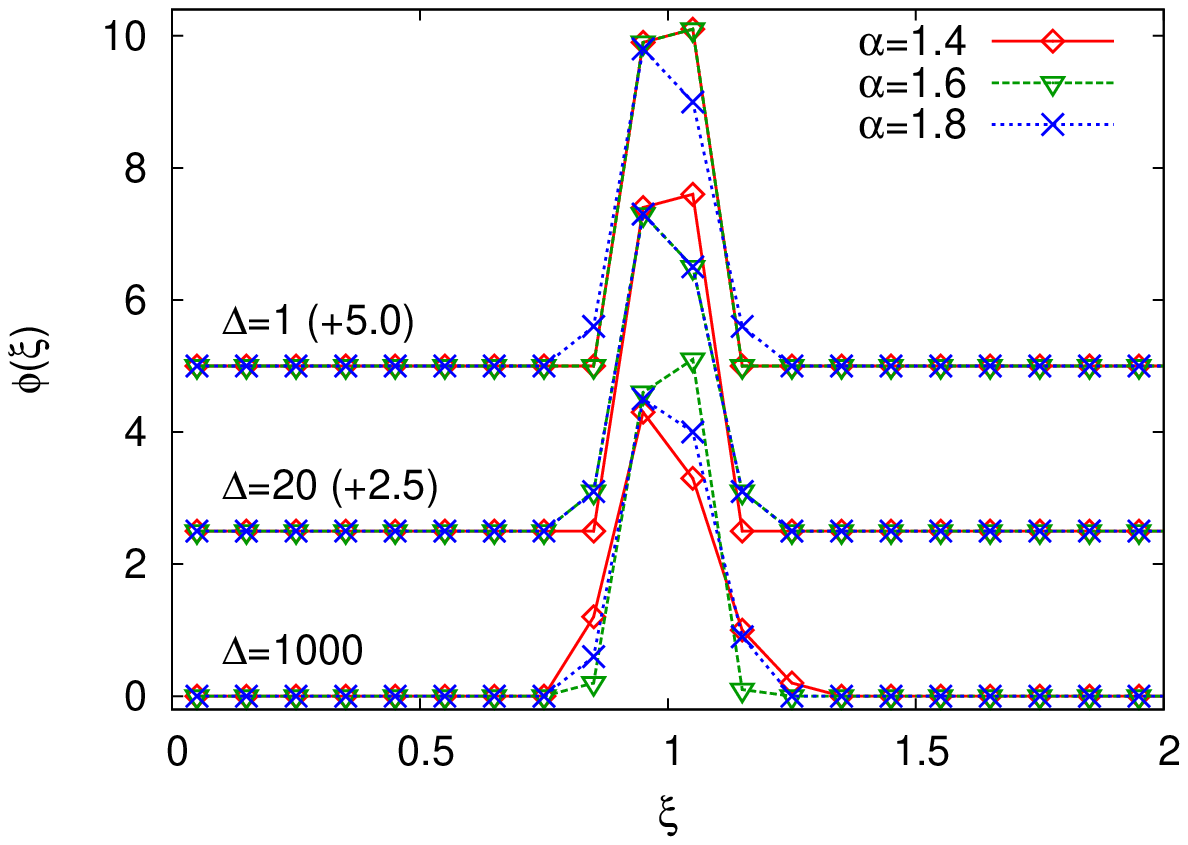}
\caption{Scatter distributions for fractional Langevin equation motion in an
harmonic potential. Left: $T=2^{11}$. Right: $T=2^{17}$.
In each graph we compare the results for three different lag times. The two
upper sets of curves were shifted by the indicated amount, for clarity.}
\label{sup2}
\end{figure*}

\begin{figure*}
\includegraphics[width=8cm]{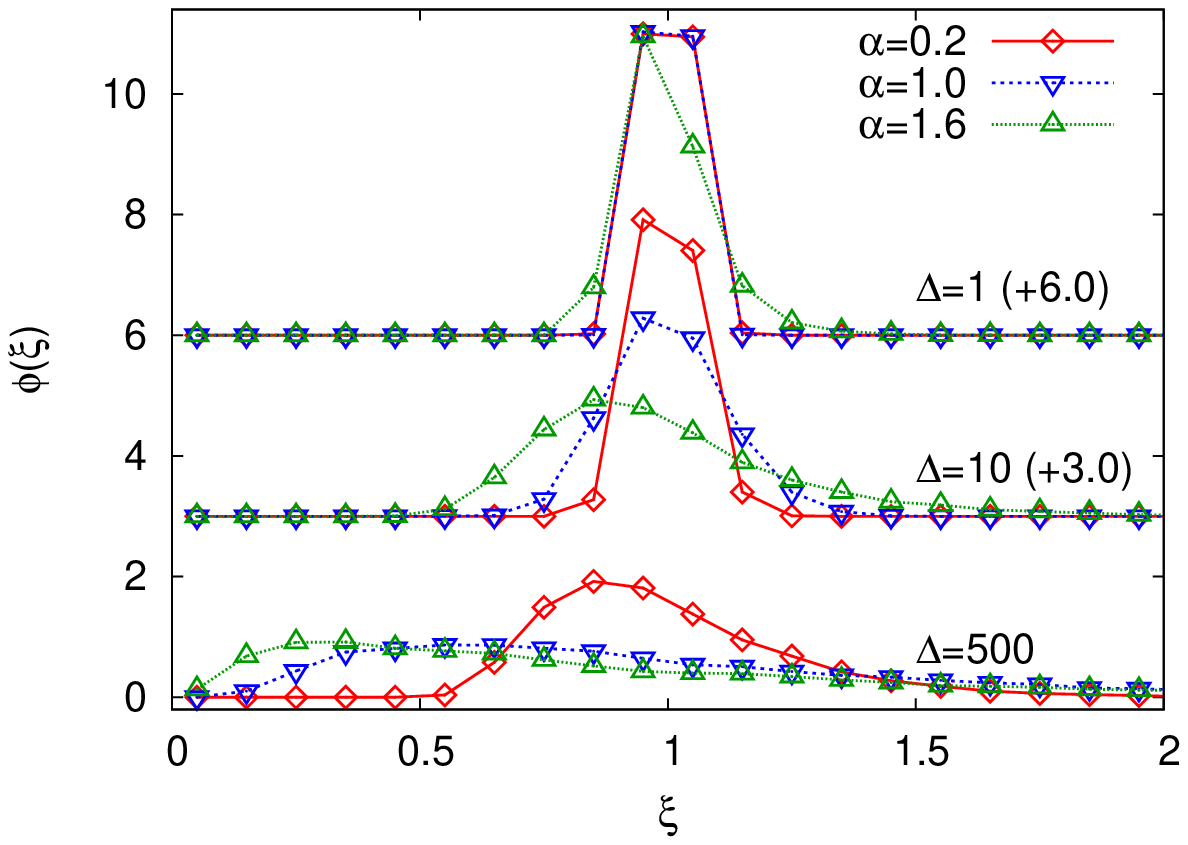}
\includegraphics[width=8cm]{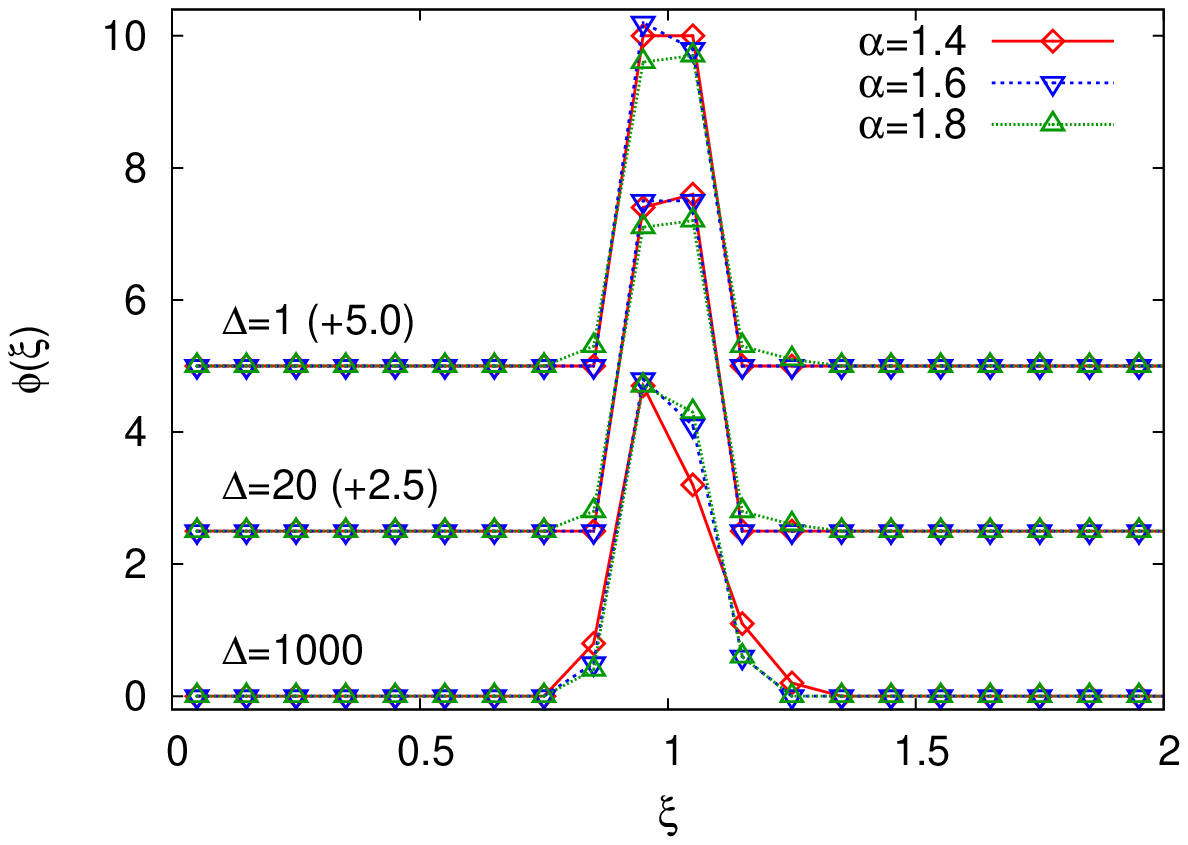}
\caption{Scatter distributions for free fractional Brownian and Langevin
equation motion. Left: $T=2^{11}$. Right: $T=2^{17}$.
In each graph we compare the results for three different lag times.
The two upper sets of curves were shifted by the indicated amount,
for clarity.}
\label{sup3}
\end{figure*}

At finite sampling time $T$ the TA MSD $\overline{\delta^2(\Delta,T)}$ is a
random variable, even for ergodic processes such as Brownian motion,
FBM, and FLE motion. In
practice, this means that $\overline{\delta^2(\Delta,T)}$ shows pronounced
trajectory-to-trajectory variations. This stochasticity of $\overline{
\delta^2}$ is measured by the scatter probability density $\phi(\xi)$, in which
the dimensionless variable $\xi$ is defined through \cite{He}
\begin{equation}
\xi=\overline{\delta^2}/\left<\overline{\delta^2}\right>.
\end{equation}
Such scatter distributions are of L{\'e}vy stable type for subdiffusive CTRW
processes with
diverging characteristic time scale \cite{He,stas2,stas3,haenggi,scatter}.
In Figs.~\ref{sup1}-\ref{sup3} we show $\phi$ for free and confined FBM and
FLE motion, with fixed bin size 0.1. Each
graph shows the distributions at three different lag times $\Delta$ and
$\alpha$ for given measurement time $T$. In each graph two sets of curves
were shifted upwards for comparison, the shift value is indicated in the graphs.

\subsection{FBM in an harmonic potential}

On the Left of Fig.~\ref{sup1} we present the distributions of $\overline{
\delta^2}$ for the data shown in Fig.~\ref{Ftamsdfbm2}. As
expected from our previous study \cite{scatter},
the distributions are centered around the ergodic value $\xi=1$
and become wider as the lag time $\Delta$ increases. The wider distribution at
longer lag time means that the single TA MSD trajectories tend to be more
erratic as $\Delta$ approaches $T$. A new finding is that at a fixed lag time
$\Delta>1$ the scatter distribution becomes broader as the motion is faster
(i.e., growing $\alpha$). This behavior is mainly due to the inherent property
of FBM itself, as the same tendency is also found without potential (see left
panel in Fig.~\ref{sup3}). This dependence on $\alpha$
is attributed to the fact that FBM is a Gaussian stationary process in which
the spatial displacement $x$ for time difference $\Delta$ is governed by
the distribution $\sim\exp(-x^2/[4K_H\Delta^{\alpha}])$. On the Right we show
the corresponding distributions when the measurement time is increased to
$T=2^{17}$. The distributions now appear insensitive to $\alpha$ and $\Delta$.
The fact that they are less sharp than the analytically predicted Gaussian is
due to the finite size effect of the binning, see below.

\subsection{FLE motion in an harmonic potential}

The left and right panels of Fig.~\ref{sup2} depict the distributions
corresponding to the parameters used in Fig.~\ref{Ftamsdglehar2},
with $T=2^{11}$ and $2^{17}$, respectively. Note that the
variation of $\alpha$ was restricted to the range [1.0, 2.0], as FLE motion
is only well defined for subdiffusion. The anomalous diffusion exponent in
this case is given by $\kappa=2-\alpha$ in terms of the scaling exponent
$\alpha$ of the fractional Gaussian noise. Generally the distributions are
bell-shaped. Notably, at fixed lag time the distribution of FLE motion tends
to be wider as the overdamped motion becomes slower (i.e., for increasing
$\alpha$), as opposed to the case of FBM. We also observe that the distributions
of FLE motion appear generally wider than those of FBM (e.g., Fig.~\ref{sup1}
Left), for the fact that the initial values of position and velocity for FLE
were chosen as the corresponding equilibrium distributions. For the case of long
measurement time (Right), the distributions again appear insensitive to
$\alpha$ and $\Delta$ for the given bin size.

\subsection{FBM \& FLE motion in free space}

To appreciate the effect of confinement on the distribution, we simulated
free FBM and FLE motion for the same parameters as in Figs.~\ref{Ftamsdfbm2}
and \ref{Ftamsdglehar2}. As shown in Fig.~\ref{sup3} in both cases, the
scatter distributions manifest features consistent with the confined cases.
Only the width of the distribution becomes narrower by the presence of the
confining potential, in particular, at longer lag times.

\subsection{Influence of bin size}

\begin{figure*}
\includegraphics[width=8cm]{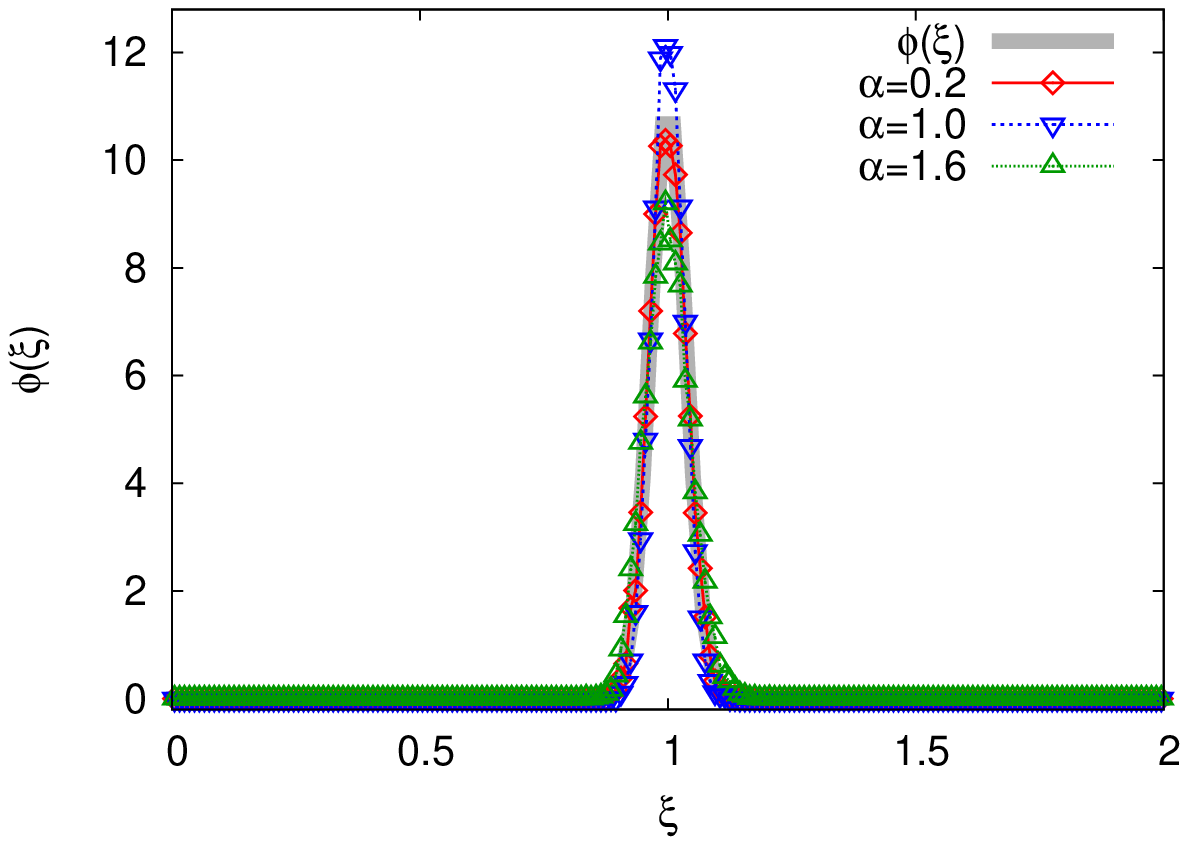}
\includegraphics[width=8cm]{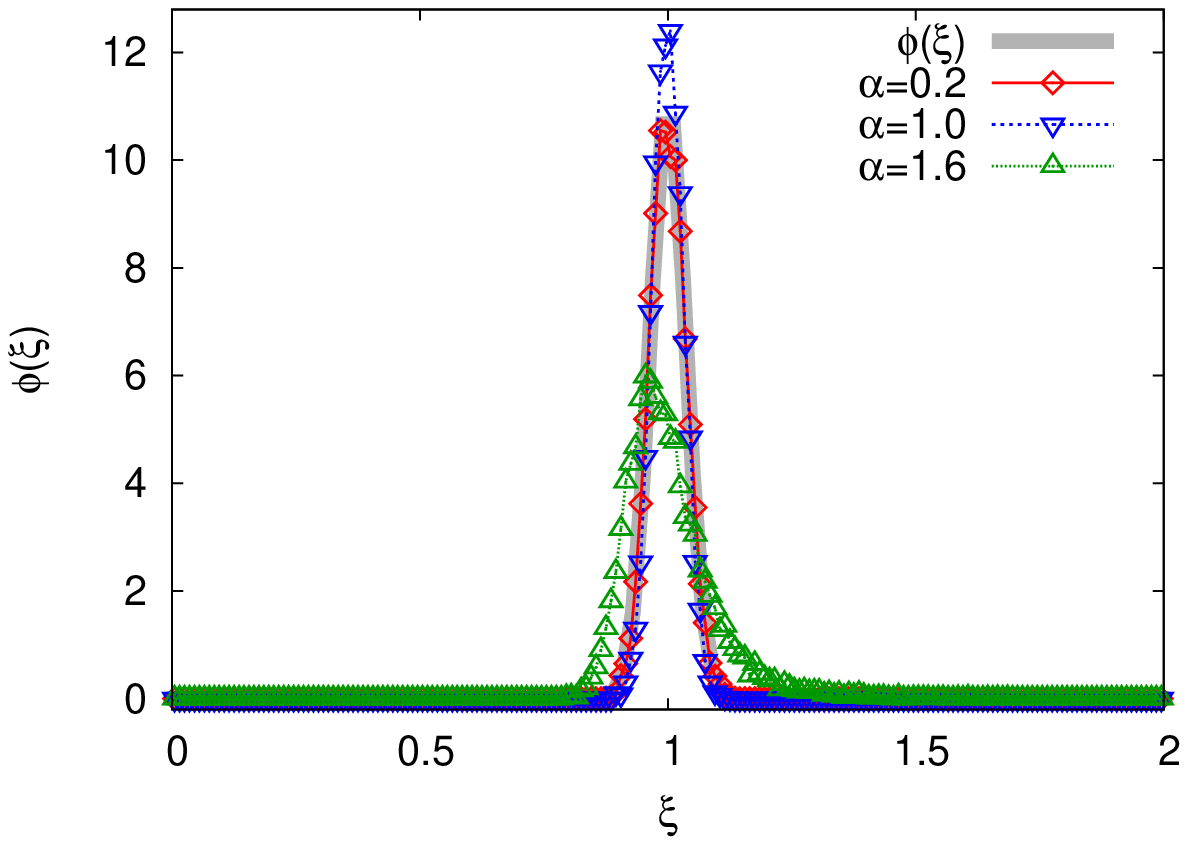}
\caption{Scatter distribution for confined (Left) and free (Right) FBM with
$T=2^{11}$, and $\Delta=1$. Data are the same as shown in Figs.~\ref{sup1}
and \ref{sup3} to the Left, but analyzed with bin size 0.01. As shown by the
grey line the measured scatter nicely agrees with the predicted Gaussian
distribution (\ref{gauss}).}
\label{sup4}
\end{figure*}

Typically experimental probing windows are limited, and meaningful quantitative
analysis requires more or less coarse binning. In the context of the scatter
plots shown here this practically means that for a bin size of 0.1 the peak
cannot exceed the value 10, for reasons of normalization, $\int_0^{\infty}
\phi(\xi)d\xi=1$. For simulations data we can arbitrarily increase the
accuracy and thus reduce bin sizes while still maintaining good statistics.
Such a result is shown in Fig.~\ref{sup4} for bin size $0.01$. For this
resolution we may compare the shape of the distribution $\phi(\xi)$ determined
from simulations with the theoretical approximation valid for short lag times
$\Delta$,
\begin{equation}
\phi(\xi)\approx\sqrt{\frac{T-\Delta}{4\pi\tau^*}}\exp\left(-\frac{(\xi-1)^2
(T-\Delta)}{4\tau^*}\right),
\label{gauss}
\end{equation}
as derived in Ref.~\cite{scatter}. Here, the scale $\tau^*$ is only introduced
to account for correct dimensions and can be taken to one [time unit], compare
Ref.~\cite{scatter}. The graph shows nice agreement with the
measured data from the simulations. Interestingly the agreement is somewhat
better in the confined case. At larger $\Delta$ the distribution becomes wider
than predicted by Eq.~(\ref{gauss}), due to the strong correlation in
successive square
displacements, $[x(t+\Delta)-x(t)]^2$ (see Ref.~\cite{scatter}), and the curves
split up for the different $\alpha$.

\section{Discussion}

Studying the representative example of ergodic FBM and FLE
motion in an external harmonic potential we demonstrated that the TA MSD
behaves significantly different from the EA MSD. Thus, naive interpretation of
single trajectory time averages based on the knowledge of the ensemble behavior
may lead to false conclusions on the physics underlying the observed motion.
This so far overlooked discrepancy is particularly relevant for the relaxation
behavior: while for the EA MSD the relaxation time can be read off directly,
the corresponding TA MSD appears to suggest a scale-free behavior.
Hence it is imperative to compare to analytical or simulations results
for the TA of the system. We note that while here we focused on an harmonic
external potential, the findings reported here also pertain to other forms of
confinement.

What is the reason for this
disagreement between EA and TA? We find that for stochastic processes converging
to a stationary state, as $\langle[x(t+\Delta)-x(t)]^2\rangle_{\mathrm{th}}$
only depends on $\Delta$, the definition of the TA MSD \eqref{tamsddef} in the
limit of long-time measurement leads to the general relation
\begin{equation}
\lim_{T\to\infty}
\overline{\delta^2(\Delta,T)}=2\langle x^2\rangle_{\mathrm{th}}[1-C_x(\Delta)],
\end{equation}
which is independent of diffusion models and details of
confinement. Here
\begin{equation}
C_x(\Delta)=\frac{\langle x(t)x(t+\Delta)\rangle_{\mathrm{th}}}{
\langle x^2\rangle_{\mathrm{th}}}
\end{equation}
is the normalized position autocorrelation
function. Therefore, the time-averaged variable $\overline{\delta^2(\Delta)}$
in fact is an indicator of the correlation of the spatial displacement, not the
TA MSD of a trajectory. For ergodic systems satisfying the Khinchin theorem
($C_x(\Delta\rightarrow\infty)=0$), the TA MSD always saturates to $\overline{
\delta^2}\to2\langle x^2\rangle_{\mathrm{th}}$, where the relaxation dynamics
reflecting the spatial correlation can be very slow although the system is
already fully relaxed, as shown in this study. Accordingly, in performing
single trajectory analysis, one should be aware of the potential pitfalls in
using $\overline{\delta^2(\Delta)}$. In contrast to non-ergodic systems
\cite{He,stas2,stas3}, the anomalous diffusion exponent $\alpha$ and the anomalous
diffusion constant $K_{\alpha}$ can be estimated from the log-log plot of
the TA MSD at short lag times. Meanwhile, physical quantities associated
with confinement such as the effective confinement size and the
relaxation time could be incorrectly deduced from the long-$\Delta$
behavior of $\overline{\delta^2(\Delta)}$.

What alternative definitions of the TA MSD could be used to mend this problem?
Instead of Eq.~\eqref{tamsddef} it would be a straightforward idea to consider
\begin{equation}
\overline{x^2(t)}=\frac{1}{t}\int_0^tx^2(t')dt'
\end{equation}
For ergodic processes with a stationary state, $\overline{x^2(t)}$ at
$t\rightarrow\infty$ equals
\begin{equation}
\langle x^2\rangle_{\mathrm{th}}=\int x^2e^{-\beta U(x)}dx\Big/\int e^{-\beta U(x)}dx.
\end{equation}
However, the time dependence
of $\overline{x^2(t)}$ is different from that of $\langle x^2(t)\rangle$.
Even for free diffusion exhibiting $\langle x^2(t)\rangle=2K_{\alpha}t^\alpha$,
the ensemble mean of $\overline{x^2(t)}$ is $\frac{2K_{\alpha}}{\alpha+1}t^
\alpha$, and thus this definition does not even work for the Brownian case.
If a dynamic variable like the MSD as function of time is concerned,
it appears that no systematic way exists for defining a TA expectation
compatible with the analogous EA.

For finite measurement time $T$ the TA MSD $\overline{\delta^2}$ shows
trajectory-to-trajectory variations, even for Brownian motion.
Consistent with previous findings \cite{scatter} for ergodic processes, the
distributions are centered around the ergodic value $\xi=1$. Importantly,
in all cases the distribution is almost independent of confinement,
except for some narrowing at long lag times.
For both FBM and FLE at long $T$ the distributions converge, while for
shorter $T$ a dependence on $\alpha$ prevails.

Concluding, the study of single trajectory averages is a non-trivial extension
of the theory of stochastic processes knowledge of which is necessary to
establish quantitative models for diffusion-limited processes in small complex
systems. The current work contributes to the development of such a theory, and
to a toolbox of diagnosis methods of the exact stochastic mechanism
underlying experimental single particle trajectories
\cite{jh,weigel,analysis,akimoto,olivier,marcin}.

\begin{acknowledgments}
We thank E. Barkai and O. Pulkkinen for stimulating discussions. Financial
support from the Academy of Finland (FiDiPro scheme) is gratefully acknowledged.
\end{acknowledgments}

\end{document}